\documentclass[
reprint,
twocolumn,
amsmath,amssymb,superscriptaddress,aps,
prl]{revtex4-1}
\usepackage[utf8]{inputenc}
\usepackage{natbib}
\usepackage{graphicx}
\usepackage{hyperref}
\usepackage{url}

\usepackage{xcolor}

\usepackage{mathtools}
\DeclarePairedDelimiter{\evdel}{\langle}{\rangle}

\begin{document}
\title{Topological universality of on-demand ride-sharing efficiency}

\author{Nora Molkenthin}
\affiliation{Chair for Network Dynamics, Institute for Theoretical Physics and Center for Advancing Electronics Dresden (cfaed), Technical University of Dresden, 01069 Dresden}
\affiliation{Network Dynamics, Max Planck Institute for Dynamics and Self-Organization (MPIDS), 37077 Göttingen, Germany}

\author{Malte Schröder}
\affiliation{Chair for Network Dynamics, Institute for Theoretical Physics and Center for Advancing Electronics Dresden (cfaed), Technical University of Dresden, 01069 Dresden}

\author{Marc Timme}
\affiliation{Chair for Network Dynamics, Institute for Theoretical Physics and Center for Advancing Electronics Dresden (cfaed), Technical University of Dresden, 01069 Dresden}
\affiliation{Network Dynamics, Max Planck Institute for Dynamics and Self-Organization (MPIDS), 37077 Göttingen, Germany}
\affiliation{Max Planck Institute for the Physics of Complex Systems (MPIPKS), 01069 Dresden, Germany}

\begin{abstract}
Ride-sharing may substantially contribute to future-compliant sustainable mobility, both in urban and rural areas. The service quality of ride-sharing fleets jointly depends on the topology of the underlying street networks, the spatio-temporal demand distributions, and the dispatching algorithms. Yet, efficiency of ride-sharing services is typically quantified by economic or ecological ad-hoc measures that do not transfer to new service regions with different characteristics. Here we derive a generic measure of ride-sharing efficiency based on the intrinsic ride-sharing dynamics that follows a universal scaling law across network topologies. We demonstrate that the same scaling holds across street networks of distinct topologies, including cities, islands and rural areas, and is insensitive to modifying request distributions and dispatching criteria. These results further our understanding of the collective dynamics of ride-sharing fleets and may enable quantitative evaluation of conditions towards increasing the feasibility of creating or transferring ride-sharing services to previously unserviced regions.
\end{abstract}

%
%

%
%
%
%
%
%
\maketitle

Transport and human mobility in particular are essential for sustainable development 
\cite{un2015_sustainbleDevelopmentGoals, carrion2012_traveTimeReliability, barbosa2018_mobility_models, macharis2018_mobilityHmanCities}.
Yet, creating and operating accessible, fair and efficient mobility systems in cities and rural areas is becoming increasingly difficult. Urbanization is projected to grow from 55\% of the 
population, about 4.2 billion people living in cities today, to 66\% of more than 10 billion people by 2050 \cite{un2014_urbanizationProspect, un2018_urbanizationProspect_keyFacts}. This densification comes with substantial social, economic and ecological challenges \cite{mcdonnell2016_urbanEcology, ramaswami2016_metaPrincipleSustainableCities, sampson2017_urbanSustainabilityInequality}. In particular, it will further increase the load on transport systems in urban areas and amplify imbalances relative to rural areas. 

\begin{figure}[h!]
    \centering
    \includegraphics[width=0.95\columnwidth]{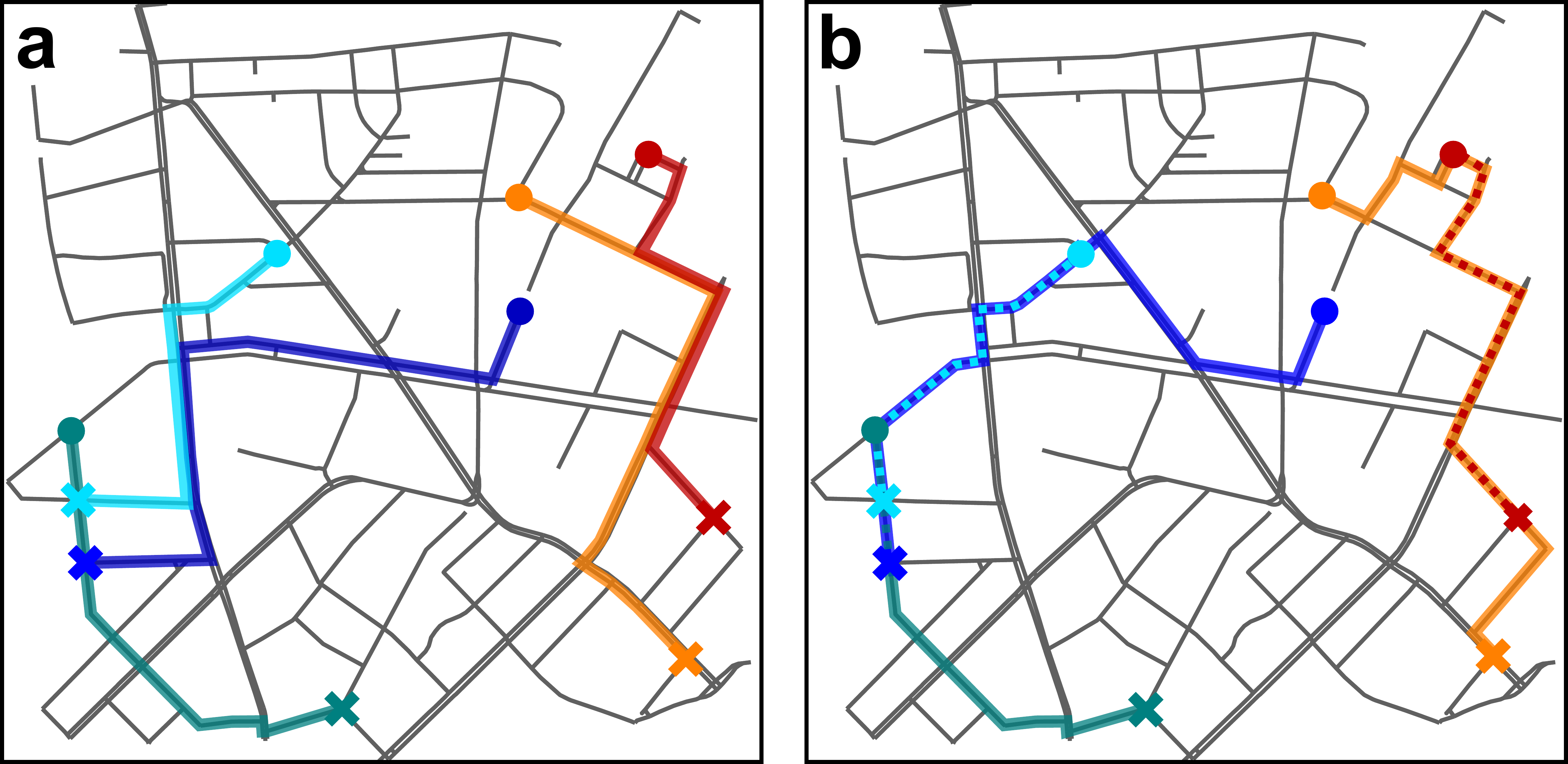}
    \caption{\textbf{Ride-sharing combines similar trips to fewer vehicles.} (a) Private car traffic and traditional ride-hailing services (e.g. taxis) serve every request individually (one color for each request, start and end points marked by disks and crosses, respectively). (b) Ride-sharing services reduce the total distance driven by combining similar requests. Here five requests are served by two vehicles, one serving three requests, one serving two, exploiting substantial overlap of the respective routes.
    }
    \label{fig:ride_sharing_sketch}
\end{figure}

Besides traditional private and public transport solutions, emerging ride-sharing services offer promising alternatives \cite{belk2014you, cohen2014ride, greenblatt2015_automated, kamargianni2016critical}. Already today, service providers such as Moia, UberPool and others \cite{moia2019, rodionova2016uberpool} operate fleets of ride-sharing vehicles and offer demand-driven transportation. These ride-sharing services combine the routes of several passengers into the same vehicle (
Fig.~\ref{fig:ride_sharing_sketch}), thereby providing options to make mobility more efficient and sustainable 
by reducing the number of cars required and the total distance driven to transport the same number of people \cite{furuhata2013ridesharing, herminghaus2019mean, santi2014_shareabilityNetworks}. Implementing such ride-sharing services requires assigning incoming requests to different transporters with suitable online-algorithms \cite{santi2014_shareabilityNetworks, sorge2017towards, alonso2017demand} and service providers require estimates for optimal fleet sizes and capacity as well as solutions to load balancing. While some of these problems have been addressed in general \cite{pavone2012_robotic, spieser2014_singapore, vazifeh2018_minimumFleetProblem, jokinen2016welfare}, many existing works focus on case studies in specific cities \cite{caulfield2009estimating, balac2015_carsharing, wright2018_public}.
In particular, the efficiency of on-demand ride-sharing services is often quantified via economic or ecological \textit{ad-hoc} measures applied to specific service conditions. The simultaneous dependence on particular cities or regions, the spatio-temporal demand patterns and the chosen dispatching algorithms make it hard to predict service efficiency and optimal parameters in new areas or under unfamiliar conditions.

In this article, we propose to quantify efficiency based on the collective nonlinear dynamics of the ride sharing fleet
by evaluating the average number of scheduled customers per vehicle as a function of the normalized system load. The resulting efficiency curves collapse to a universal 
scaling function across various graph theoretical model topologies as well as a wide range of real world street network topologies, including cities of different sizes and densities, rural areas and islands. A single topological factor measures the difficulty of implementing an efficient ride-sharing system and quantifies the impact of the specific network topology and request distribution. The universal scaling uncovered may be relevant to the large scale implementation of ride-sharing, as it supports prediction of ride-sharing efficiency not 
only in newly serviced cities 
but also its adaptation to suburban or rural areas with qualitatively different demand conditions.

\section*{Results}

Ride-sharing fleets 
operate similar to standard ride-hailing services such as taxi fleets but with several passengers sharing the same vehicle (bus). A customers requesting a ride is served by one of $B$ buses that 
may pick up and deliver additional customers who share a similar route, compare Fig.~\ref{fig:ride_sharing_sketch}. The collective dynamics of any such ride-sharing service crucially depends on three factors: (i) the locations of potential pick-up and drop-off stops and the street network connecting them, (ii) the demand distribution, i.e., the distribution of origin and destination of all requests in space and time, and (iii) the dispatcher algorithm that assigns the incoming requests to a specific bus and plans the routes of all buses. Here, we focus on the impact of the topology of the street network on the efficiency of ride-sharing, evaluating ride-sharing dynamics on various empirical and model street networks. The demand distribution naturally enters as it modifies the effective topology created by the vehicles' driving pattern. See Methods for details and the Supplementary Information for examples for different request distributions and dispatcher algorithms.

\subsection*{Efficiency of ride-sharing}
Efficient ride-sharing requires a sufficient density (in space and time) of requests. At low request rate $\lambda$, sharing rides would require customers to wait for other similar requests and impose long delays. As $\lambda$ increases, more rides are requested in a given time and the likelihood that a ride can be shared with a similar request increases \cite{tachet2017scaling}. At the same time, for a given number of buses $B$, more trips \emph{need} to be shared to serve all requests. We measure the load on a system by the normalized request rate, $x = \frac{\evdel{l}}{vB} \, \lambda$, where $\evdel{l}$  is the average trip length per customer and $v$ the characteristic 
bus driving velocity (see Methods for more details). A longer average trip length $\evdel{l}$ implies that buses are busy with individual requests longer and the load is higher. Increasing the driving velocity $v$ or the number $B$ of buses reduces the load per bus. If $x < 1$, all requests can be served one by one (e.g. by taxis), if $x > 1$ ride-sharing becomes necessary.

\begin{figure}[h!]
        \centering
	    \includegraphics[width=0.95\columnwidth]{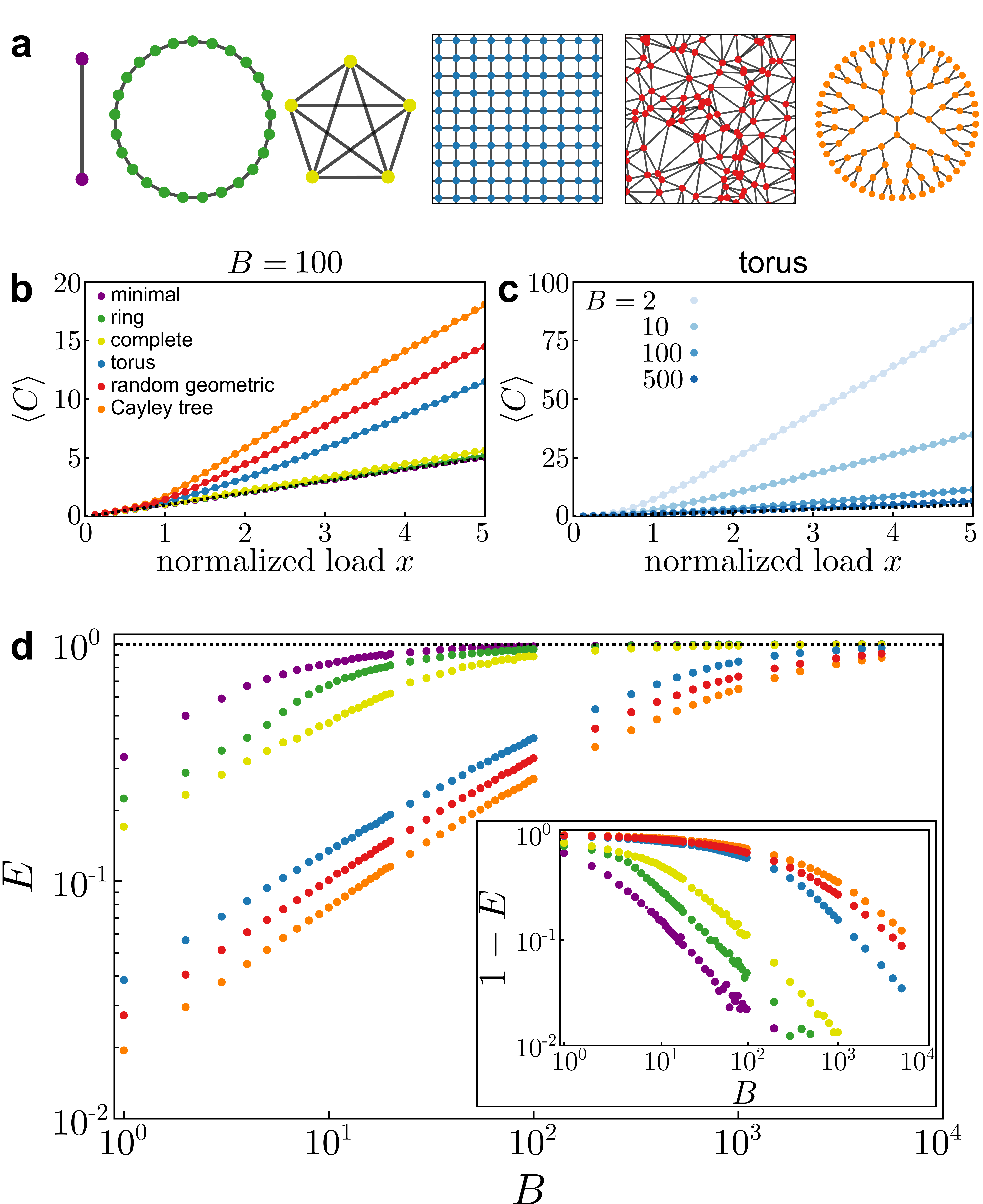}
        \caption{\textbf{The scaling of the number of scheduled customers measures ride-sharing efficiency.} (a) Model networks with qualitatively different topologies: a minimal graph ($N=2$, purple), a cycle graph (ring, $N=25$, green), a complete graph ($N=5$, yellow), a square lattice with periodic boundaries (torus, $N=100$, blue), a random geometric network ($N=100$, red) and a Cayley tree with degree $3$ ($N=94$, orange), see Methods for details.
        (b) The average number of scheduled customers $\evdel{C}$ grows linearly with the normalized request rate $x$ in the ride-sharing regime ($x > 1$). Ride-sharing is easier in networks with few distinct shortest paths (e.g. ring, green) and $\evdel{C}$ is closer to the optimal service scaling $\evdel{C} = x$ (black dashed line). The colored lines indicate the expected number of customers from the observed waiting and driving times, Eq.~\eqref{eq:c}.  
        (c) The scaling of $\evdel{C}$ converges to the optimal scaling as the number of buses is increased. At constant load $x$ the number of requests increases proportionally to the number of buses, also increasing the number of similar trips that can be shared efficiently. See also Supplementary Figure S2. 
        (d) The difference to the optimal scaling defines the efficiency $E$ [Eq.~\ref{eq:eff}, evaluated at $x = 7.5$]. The quantitative value of the efficiency varies strongly across the different topologies while the qualitative behavior is similar . 
        }
        \label{fig:eff}
\end{figure}

What are suitable observables to quantify ride-sharing efficiency? Instead of focusing on specific resources, such as fuel consumption or monetary cost, we here evaluate efficiency based on the intrinsic fleet dynamics.
At any time, each bus of a ride-sharing service has a number $C$ of customers it is scheduled to serve, including passengers already on the bus as well as customers planned to be picked up in the future. As the load on the system increases, per bus more customers are scheduled and served. 
Fig.~\ref{fig:eff} illustrates the scaling of the average number of scheduled customers $\evdel{C}$ for various model networks. 
If the number of scheduled customers exactly reflects the load on the system, i.e. per bus the fleet serves $\evdel{C} = x$ customers at any given time, the system is operating at optimal efficiency. Consider, for example, the onset of ride-sharing, $x = 1$: in the limit of perfectly efficient service, the buses on average have exactly $\evdel{C} = 1$ customer scheduled at each time. Otherwise, when the buses serve individual requests less efficiently with a lower rate, they have $\evdel{C} > 1$ customers scheduled. 
The deviation from the ideal scaling $\evdel{C} = x$ naturally measures the efficiency 
\begin{equation}
    E = \lim_{x \rightarrow \infty} \left(\frac{\evdel{C}}{x}\right)^{-1} \, \label{eq:eff}
\end{equation}
in terms of the intrinsic dynamics of the ride-sharing system. 
In networks where shortest paths between different pairs of nodes coincide and rides can be easily shared (e.g. on ideal ring networks), the optimal efficiency $E = 1$ is easier to reach than in networks with many distinct, non-overlapping shortest paths (e.g. trees), see Fig.~\ref{fig:eff}(d). Measuring the deviation from the optimal scaling at a given load $x$ defines the susceptibility of the system to changes of the load, \mbox{$\chi = \left(\frac{\mathrm{d}\,\evdel{C}}{\mathrm{d}\,x}\right)^{-1}$}, describing the efficiency with which the system handles \emph{additional} requests. In the limit of high load, the susceptibility becomes identical to the efficiency,  $E = \lim_{x \rightarrow \infty} \chi$, due to the linear scaling of $\evdel{C}$ for large $x$ [compare  Fig.~\ref{fig:eff}(b,c) and Supplementary Information].

\subsection*{Topological universality}
\begin{figure}[h!]
        \centering
	\includegraphics[width=0.9\columnwidth]{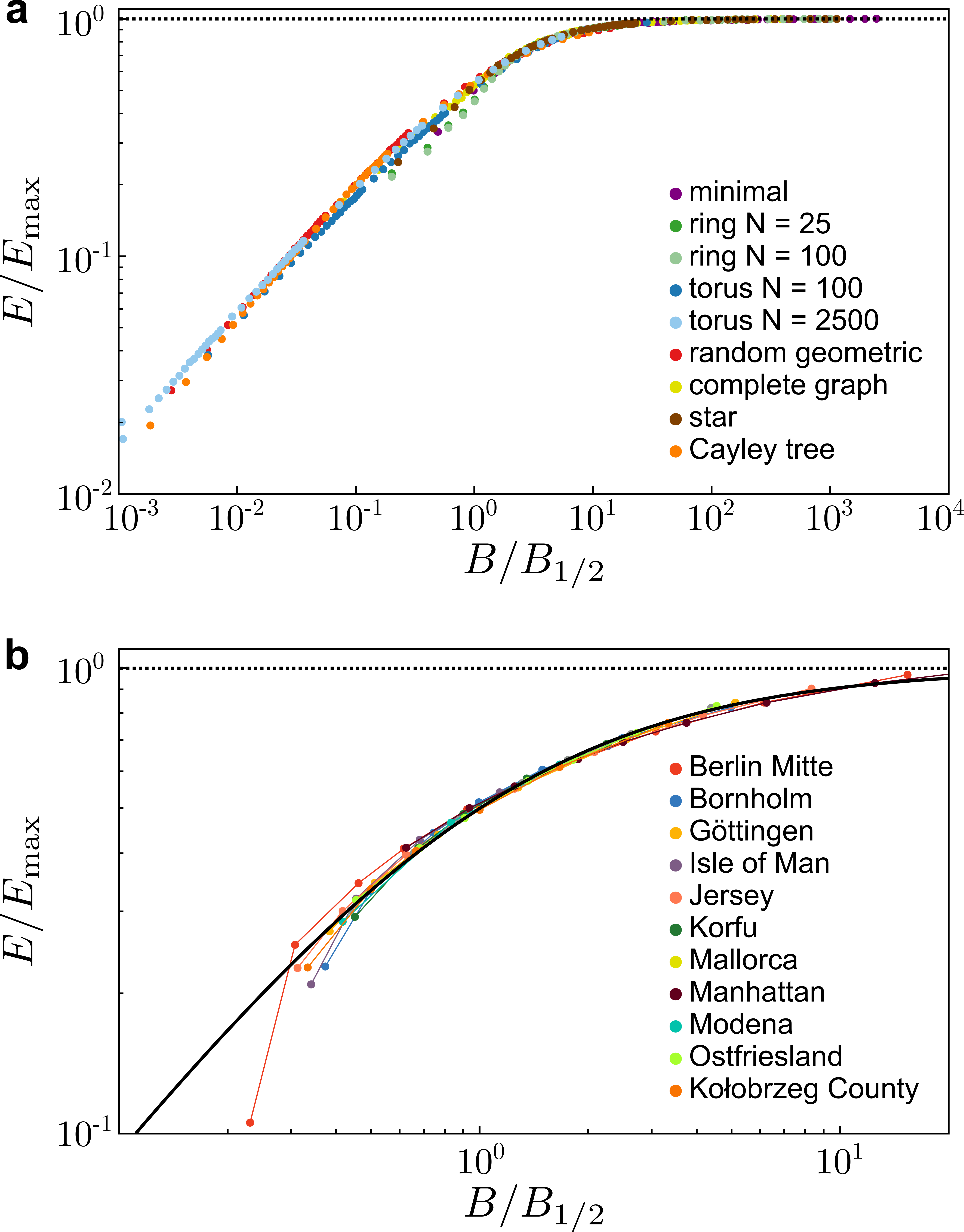}
        \caption{\textbf{Ride-sharing efficiency is universal across qualitatively different real and model street networks.} 
        (a) The ride-sharing efficiency $E$ [Eq.~(\ref{eq:eff}), evaluated at $x=7.5$] in qualitatively different model networks (compare Fig.~\ref{fig:eff} and Methods) collapses to a universal efficiency function $f(B / B_{1/2})$ [Eq.~(\ref{eq:universal})]. $E_\mathrm{max} = 1$ for all networks due to the choice of dispatcher.  
        (b) The ride-sharing efficiency $E$ (evaluated at $x=2.5$) in empirical street networks similarly collapse to a universal scaling function $f(B / B_{1/2})$. The black line indicates the theoretical prediction for large $B$ [Eq.~(\ref{eq:derived_universal})].
        }
        \label{fig:universal}
\end{figure}

The similarity between the efficiency curves in Fig.~\ref{fig:eff}(d) suggests a universal scaling of efficiency with the number of buses. In fact, extensive numerical simulations indicate that all efficiency curves collapse to
\begin{equation}
    E = E_\mathrm{max} \, f\left(\frac{B}{B_{1/2}}\right) \,.\label{eq:universal}
\end{equation}
$f\left(\cdot\right)$ is a universal efficiency function and the influence of the network structure can be summarized in a single topological scaling factor $B_{1/2}$ [Fig.~\ref{fig:universal}(a)]. This universal scaling law holds across various model topologies as well as a range of qualitatively different empirical street networks of cities of different sizes and densities, rural areas, and islands, see Fig.~\ref{fig:universal}(b). Moreover, it is insensitive against varying the dispatching algorithm (Supplementary Figure S3 and S4) and holds across a range of request distributions with uncorrelated and correlated as well as symmetric and asymmetric origin-destination pairs  (Supplementary Figure S5).


The scaling factor $B_{1/2}$ denotes the number of buses required to reach half the maximum possible efficiency $E_\mathrm{max}$ and depends on the network topology and request distribution. The maximum efficiency $E_\mathrm{max}$ strongly depends on the dispatcher algorithm. If the dispatcher does not delay any customer in the perfect service limit, the efficiency approaches $E_\mathrm{max} = 1$.

\subsection*{Scaling of ride-sharing efficiency}
How does the scaling function $f(\cdot)$ relate to the observables of the ride sharing dynamics? To address this question, we derive the asymptotic scaling for large numbers of buses and requests, enabling us to estimate the scaling factor $B_{1/2}$.

\begin{figure}[t]
        \centering
	\includegraphics[width=0.75 \columnwidth]{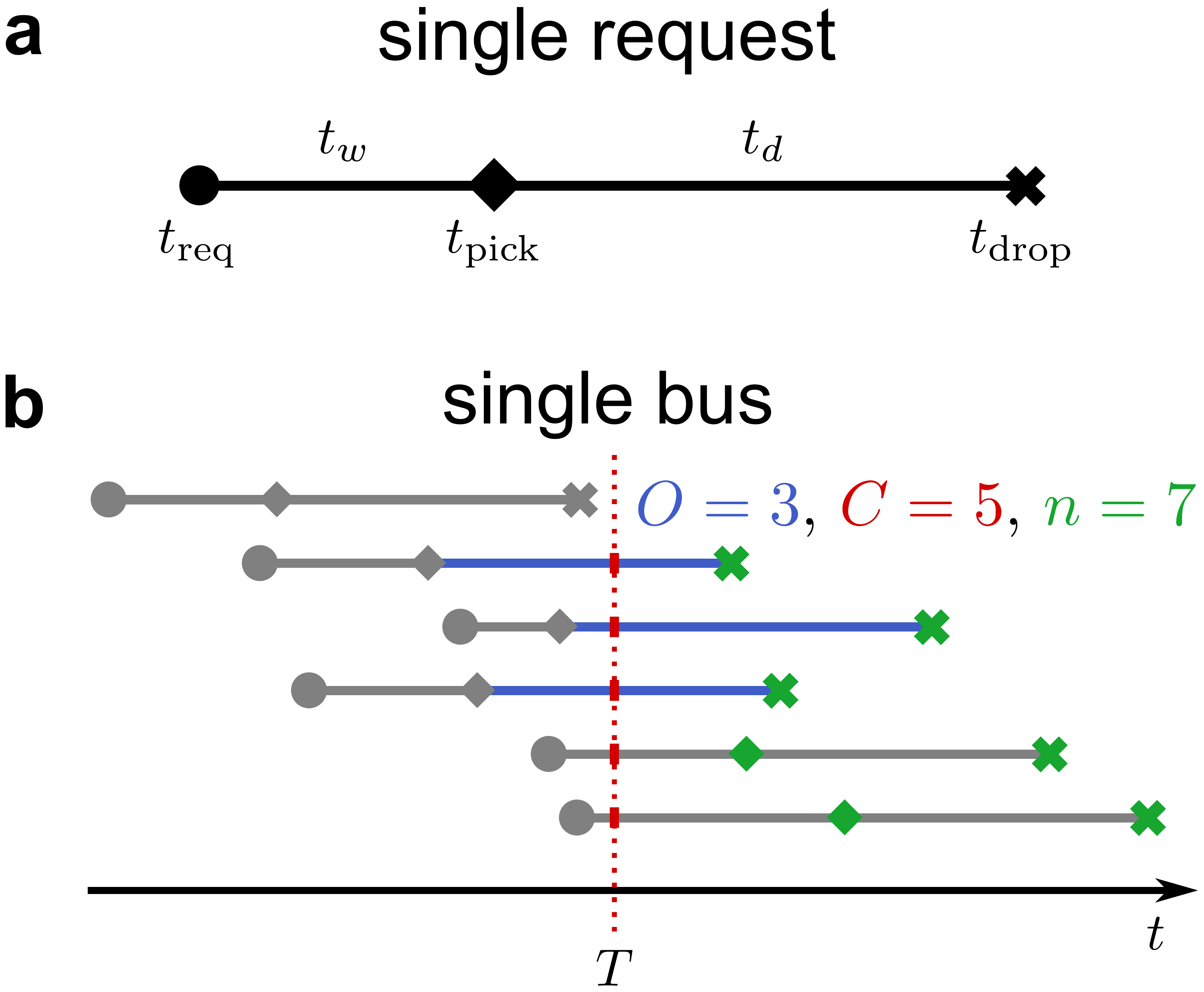}
        \caption{\textbf{Waiting and driving time of requests determine number of scheduled customers per bus.}
        a) A request is made at time $t_\mathrm{req}$, picked up at time $t_\mathrm{pick}$ and delivered at their destination at time $t_\mathrm{drop}$. They thus spend a time $t_w = t_\mathrm{pick} - t_\mathrm{req}$ waiting to be picked up and a time $t_d = t_\mathrm{drop} - t_\mathrm{pick}$ in transit on the bus.
        b) The route of each bus consists of a number of overlapping customer trips. At time $t = T$ there are $O$ passengers on board, which is equivalent to the number of overlapping drive times (blue). The number of scheduled customers $C$ equals the total number of overlapping service times (red). The number of scheduled stops $n$ is the total number of pick-up and drop-off points scheduled, but not yet served, at time $T$ (green). On average, these quantities are directly related to the waiting and driving time [see Eq.~\eqref{eq:c}].
        }
        \label{fig:time}
\end{figure}

The average number of scheduled customers is directly related to the average waiting time $\evdel{t_w}$ until pickup and driving time $\evdel{t_d}$ between pick-up drop-off of an individual customer [Fig.~\ref{fig:time}(a)]. During the time interval between a customer making a request and that customer arriving at their destination, a bus is assigned new requests with an average rate $\lambda / B$. Over the average service time of a customer, $\evdel{t_s} = \evdel{t_w} + \evdel{t_d}$, the bus thus schedules on average $\lambda \evdel{t_s} / B$ new requests. At the expected time this average customer leaves the bus, 
only those new requests are expected to still be scheduled, while the older ones are expected to have been delivered earlier,
such that the average number of scheduled customers is

\begin{eqnarray}
 \evdel{C} = \lambda \, \frac{\evdel{t_s}}{B} = \frac{v x}{\evdel{l}} \left(\evdel{t_d} + \evdel{t_w}\right) \,.\label{eq:c}
\end{eqnarray}
This argument is illustrated in Fig.~\ref{fig:time}(b). Similar arguments relate the average number $\evdel{O}$ of customers currently on a bus (its occupancy) and the average number $\evdel{n}$ of planned stops to the waiting and driving time, $\evdel{O} = \frac{v x}{\evdel{l}} \evdel{t_d}$ and $\evdel{n} = \frac{v x}{\evdel{l}} \left(\evdel{t_d} + 2 \, \evdel{t_w}\right)$ [compare Fig.~\ref{fig:time}(b)].

To derive the scaling of the efficiency curve, we first consider the scaling of $\evdel{C}$ close to the perfect service limit. This means we consider large $B \rightarrow \infty$ for perfect service [compare Fig.~\ref{fig:eff}(c)] and large $x \gg 1$ as in the definition of ride-sharing efficiency Eq.~(\ref{eq:eff}). For an efficient ride-sharing dispatcher algorithm, the delay due to detours disappears in the perfect service limit, such that 
\begin{equation}
    \evdel{t_d} \sim \frac{\evdel{l}}{v} \propto B^0
    \label{eqn:td_scaling}
\end{equation} to leading order in $B$.

The waiting time is determined by the number of buses going directly from the origin to the destination of a request. When there are sufficiently many buses in the network, multiple buses drive along each shortest path in the network. Consequently, the waiting time decays to zero as the number $B$ of buses becomes large, scaling proportional to $B^{-1}$, since twice as many buses means a bus going in the right direction comes by twice as often. We express the waiting time in terms of the natural time scale $\tau=\frac{\evdel{l}}{v}$ in the system and a proportionality factor $\gamma$ reflecting a characteristic number of buses at which the average waiting time equals the system-intrinsic time scale $\tau$. We thus obtain 
\begin{equation}
    \evdel{t_w} \sim \gamma \tau B^{-1} \propto B^{-1}
    \label{eqn:tw_scaling}
\end{equation}
 for large $B$. Substituting Eq.~\eqref{eqn:tw_scaling} and \eqref{eqn:td_scaling} into Eq.~\eqref{eq:c} yields
 \begin{equation}
    \evdel{C} \sim \frac{v x}{\evdel{l}} \left(\frac{\evdel{l}}{v} + \gamma \frac{\evdel{l}}{v} B^{-1}\right) = x \left(1 + \frac{\gamma}{B}\right) \,, 
\end{equation}
and with Eq.~\eqref{eq:eff} the universal scaling law 
\begin{equation}
    E = E_\mathrm{max} \, f\left(\frac{B}{\gamma}\right) \,, \label{eq:derived_universal}
\end{equation}
for the ride-sharing efficiency [compare Eq.~\eqref{eq:universal}], where the asymptotic scaling function is $f(z)=1/(1+z^{-1})$ as $z \rightarrow \infty$ and we directly identify $\gamma = B_{1/2}$ as the number of buses required to reach half efficiency (see Supplementary Figure S1).

Universality implies that there is a single scaling function $f(\cdot)$ and scaling factor $B_{1/2}$ valid across the entire range of $B$, despite the above derivation relying on the asymptotic scaling as $B \rightarrow \infty$. In particular, the scaling factor $B_{1/2}$ analytically follows from this asymptotic scaling for a range of model networks. We remark that the above derivation of Eq.~\eqref{eq:derived_universal} includes the effect of the request distribution on the scaling factor by calculating the average trip length $\evdel{l}$ with respect to the request distribution.

\subsection*{Distinctness of shortest paths controls scaling factor}
These results raise the question which topological properties most strongly influence the efficiency function through its scaling factor $B_{1/2}$.

\begin{figure}[h!]
        \centering
	\includegraphics[width=0.8\columnwidth]{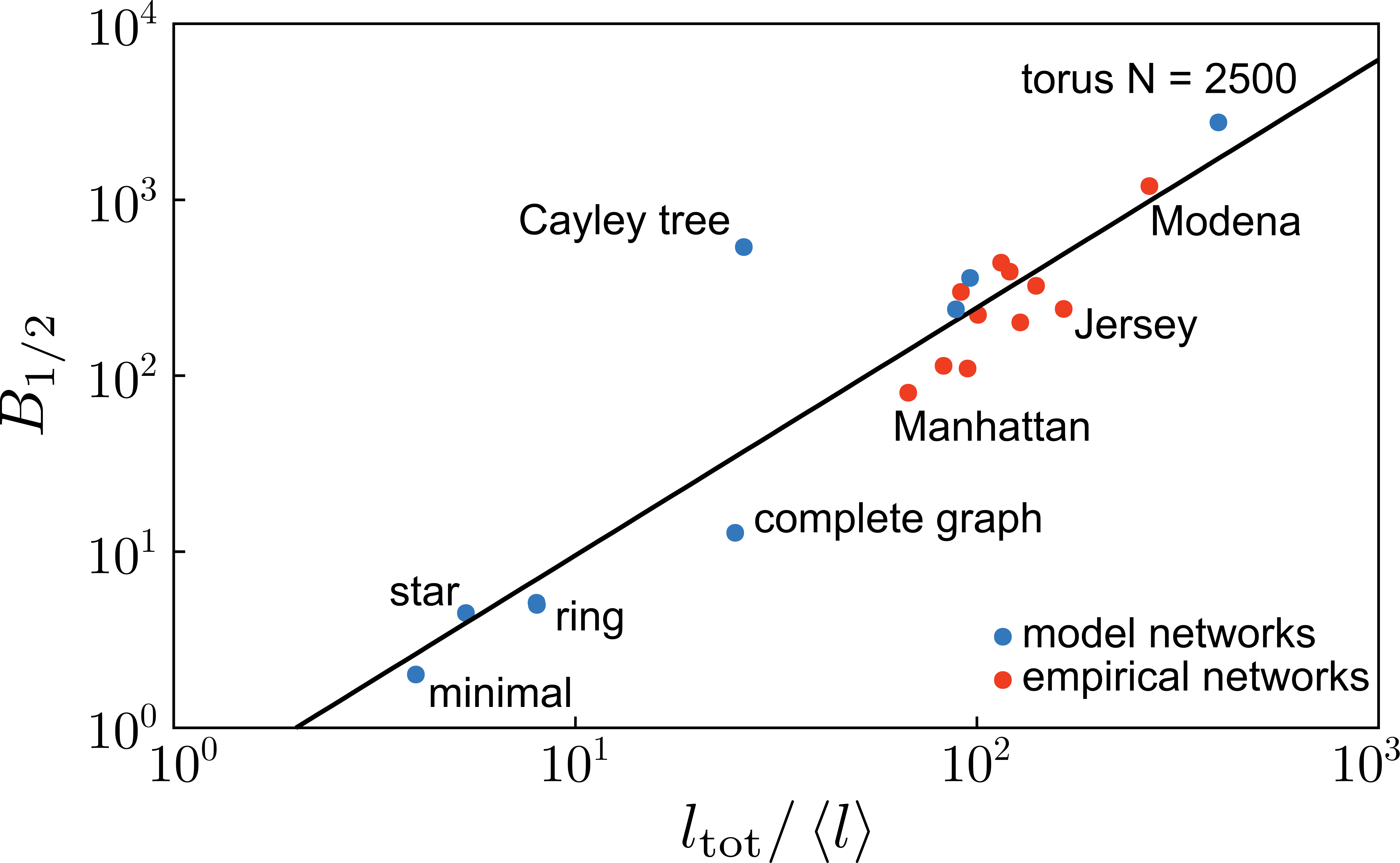}
        \caption{\textbf{Distinct shortest paths make efficient ride-sharing more difficult.} 
        The scaling factor $B_{1/2}$, denoting the number of buses required to reach half the maximum possible efficiency, increases as shortest paths become more distinct, quantified by the ratio of the total length $l_\mathrm{tot}$ of all links in the network to the average shortest path length $\evdel{l}$. The black line indicating the best linear fit is a guide to the eye. The Cayley tree and complete graph represent the graph theoretically most extreme network topologies, far different from real street networks. 
        See Methods for details on the simulation and Supplemental Information for the numerical values of the scaling factor $B_{1/2}$ and $E_\mathrm{max}$ of the networks.
        }
        \label{fig:overlap}
\end{figure}

The extent to which rides can be shared in the network topology is intuitively measured by the overlap or similarity of shortest paths in the network. Conversely, many distinct, non-overlapping shortest paths imply that efficient ride-sharing is difficult, quantified by the scaling factor $B_{1/2}$. We measure this distinctness of shortest paths as the ratio $\ell = l_\mathrm{tot}/\evdel{l}$ between the total length $l_\mathrm{tot}$ of all links in the network and the average shortest path length $\evdel{l}$. If $\ell$ is small, a typical shortest path use many edges in the network, indicating high overlap between different shortest paths (e.g. minimal and ring networks). This allows rides to be shared easily. If $\ell$ is large, shortest paths use only a small fraction of links and most shortest paths are distinct from one another or only share few edges (e.g. torus network). Consequently, trips are harder to share and we expect $B_{1/2}$ to be larger. 
Figure~\ref{fig:overlap} indeed shows a strong dependence of $B_{1/2}$ on the distinctness $\ell$ of shortest paths across all real and model networks.

\section*{Conclusion}
Understanding and quantifying the dynamics of on-demand ride-sharing systems is essential to plan and operate ride-sharing fleets to provide a viable alternative to private cars \cite{kamargianni2016critical, macharis2018_mobilityHmanCities}. In this article, we have introduced a general, quantitative measure of ride-sharing efficiency based on system-intrinsic dynamics, enabling researchers and service providers to quantify the total efficiency of the system (absolute efficiency) as well as its response to changes in the demand (susceptibility).

The proposed efficiency measure exhibits a universal scaling law across different network topologies that is insensitive to varying demand distributions and dispatching algorithms. This scaling law quantitatively characterizes the influence of a network's size, topology and request density on the efficiency of ride-sharing. Two parameters, $E_\mathrm{max}$ and $B_{1/2}$, of a universal efficiency curve summarize the effects of topology, request distribution and dispatcher algorithm. The maximum possible efficiency $E_\mathrm{max}$ strongly depends on the choice of dispatching algorithm, while the scaling factor $B_{1/2}$ quantifies the number of buses and requests required for efficient ride-sharing to be possible for the given network topology and request distribution.

Previous studies found universal scaling of shareability properties \cite{santi2014_shareabilityNetworks, tachet2017scaling}, allowing comparison of the theoretical potential of ride-sharing across different cities. Our results now enable prediction of the actual efficiency of ride-sharing systems operating under a wide variety of different conditions. 
However, by its very nature, the universality cannot hold across arbitrary services and conditions. For instance, if the ride-sharing fleet itself generates the majority of the traffic in a city, the traffic congestion and thereby the characteristic driving velocity $v$ will explicitly depend on the request rate $\lambda$ and the number of vehicles $B$. As a consequence, the relations in Eqs.~\eqref{eq:c}-\eqref{eqn:tw_scaling} do not imply the scaling Eq.~\eqref{eq:derived_universal} because $B$ couples to $\lambda$ and thus $x$. Moreover, our derivation of the scaling assumes an asymptotically constant driving time and a waiting time scaling as $B^{-1}$ for large $B$, a reasonable assumption for most dispatchers. However, the same asymptotic universality is not guaranteed to hold for hypothetical dispatchers with a different scaling. 

These constraints notwithstanding, our results may help to transfer insights from individual ride-sharing systems or case studies in specific cities to other cities and rural areas of different sizes, densities and with qualitatively different street networks and demand distributions. This not only enables better planning of ride-sharing fleets in previously unserviced areas, but also the use of data-driven automated methods to select suitable dispatcher algorithms and service parameters by making data from different settings comparable \cite{vazifeh2018_minimumFleetProblem}. Furthermore, the dependence of the efficiency on the network topology (e.g Fig.~\ref{fig:overlap}) suggests a way to systematically optimize ride-sharing services topologically, for example by constraining stop locations \cite{li2018modeling, moia2019} and allowing routes that form an effective network more suited for efficient ride-sharing.


\scriptsize

\section*{Methods}

\subsection*{Dynamics of ride-sharing}
The ride-sharing dynamics are a stochastic process on a street network with $N$ nodes (intersections or places of interest) and $M$ weighted and directed links (streets), where the weight $l(i,j)$ of the link $(i,j)$ denotes the distance between the connected nodes $i$ and $j$. It may equally describe the time required to travel on the street, thus including effects of heterogeneous velocities due to congestion or speed limits. On this network, we consider a fixed number $B$ of buses with infinite passenger capacity and characteristic velocity $v$.

Each bus $b$ has a planned route consisting of $n_b(t)$ scheduled stops to pick up or deliver customers. This route also defines our main observables: the occupancy $O_{b}(t)$, i.e., the number of passengers on bus $b$ at time $t$, the number of scheduled customers $C_{b}(t)$ and the number of scheduled stops $n_{b}(t)$ [compare Fig.~\ref{fig:time}(b)]. Note that, if a bus has to pick up or deliver several customers at the same node $i$, this node $i$ appears multiple times in the route, once for each customer.

We model the requests by customers as a Poisson process with constant rate $\lambda = 1/\evdel{\Delta t}$, i.e. with time $\Delta t$ between requests distributed exponentially with mean $\evdel{\Delta t} = 1 / \lambda$. When a customer makes a request at time $t_\mathrm{req}$, the dispatcher algorithm assigns a bus $b$ to include the new request in its route as well as the details of the insertion of the required stops into the route of the bus. Over time, the bus will make all scheduled stops and eventually pick up the customer after a waiting time $t_w$ and deliver them to their destination after an additional driving time $t_d$ [see also Fig.~\ref{fig:time}(a)].\\

The parameters (network topology, request distribution and dispatcher algorithm) used in the simulations presented in the manuscript are described below. Results for additional parameters and further details are given in the Supplementary Information.

\subsection*{Normalized load $x$}
To be able to compare the dynamics across different street networks, request patterns or different numbers of transporters and request rates, we define a normalized load $x$ by considering how many requests a single taxi ($B=1$ bus with capacity $O_\mathrm{max} = 1$) can handle. The taxi needs on average a time $\evdel{t_w^{(\text{taxi})}}+\evdel{t_d^{(\text{taxi})}}$ to pick up a customer and drop them off at their destination. In the optimal case, the pickup time vanishes and the taxi can serve requests up to a rate $\lambda^{(\mathrm{taxi})}_\mathrm{max} = 1/\evdel{t_d^{(\text{taxi})}} = v / \evdel{l}$ given by the average drive time without any detours. From this consideration, we define the normalized load $x$ as the effective request rate $\lambda / B$ per bus relative to this maximal rate
\begin{equation}
 x = \frac{\evdel{l}}{v B} \, \lambda \,, 
 \label{eq.x}
\end{equation}
where the average shortest path length 
$\evdel{l}$ is taken with respect to the request distribution in the network. Loads $x < 1$ can be handled by a taxi service. For loads $x > 1$, ride-sharing is required to serve all requests. For example, $x = 3$ means that each bus has to serve three times as many requests as a taxi could handle, something that would be impossible without ride-sharing.

\subsection*{Event based simulation}
We simulate the ride-sharing dynamics with an event-based approach. Events are divided into bus events, where a bus collects or delivers a customer, and request events, where a new request is made. 
Each bus $b\in \{1,...,B\}$ is described by its current location and its route $R_{b}(t) = \left(i_{1}^{(b)},...,i_{n}^{(b)}\right)$ at time $t$ as an ordered list of nodes $i_{k}^{(b)}$ that the bus is scheduled to stop at to pick up or deliver a customer.\\

\textit{Request event: }For a request event at time $t_\mathrm{eq}(k)$ we randomly choose an origin $i(k)$ and a destination $j(k)$ according to a given request distribution $P(i,j)$ (see below). We then find all possible offers of every bus, based on the origin and destination of the request and the current positions and scheduled routes of the buses. We assign the request to the bus $b(k)$ with the ``best'' offer according to the dispatcher algorithm minimizing, for example, the arrival time $t_\mathrm{drop}$ (see below). This bus inserts the two additional stops to pick up and deliver the customer into its route according to the dispatcher decision. Finally, a new request is generated at time $t_\mathrm{req}(k+1) = t_\mathrm{req}(k) + \Delta t$ where $\Delta t$ is distributed exponentially with rate $1/\left<\Delta t\right>$, meaning the requests follow a Poisson process in time.\\

\textit{Bus event: }For a bus event we update the position of the bus. If the bus reaches a scheduled stop, we simply adjust the occupancy of the bus when collecting or delivering a customer (we assume passengers enter and exit buses instantaneously). When a customer is delivered we record the statistics of the trip, such as request time $t_\mathrm{req}$, pickup time $t_\mathrm{pick}$ and drop off time $t_\mathrm{drop}$. Finally, we determine the next event for this bus based on its scheduled route and the bus drives to its next destination. Otherwise, if there are no more scheduled stops, the bus remains at its current location and waits idly until a new request is assigned to it.\\

We start all simulations in the empty state without any requests, all buses are initially idle and uniformly randomly distributed over all nodes in the network. We let the system equilibrate for some time, typically $100$ requests per bus, before measuring for at least $1000$ requests per bus to determine the results reported in the figures.

\subsubsection*{Dispatching algorithms}
For simulations on the model networks, we use a dispatcher algorithm minimizing the arrival time $t_\mathrm{drop}$ of each new customers without delaying previous requests. If multiple transporters offer the same arrival time (by coincidence), this dispatching algorithm chooses the bus with the smallest drive time $t_d = t_\mathrm{drop} - t_\mathrm{pick}$ and then (if still multiple transporters are possible) the bus with the currently largest number of passengers (to potentially let another bus become idle and more effectively serve other requests).

In the empirical networks, the shortest paths are often unique and allowing no detours would result in unrealistic behavior of the buses. We thus employ an algorithm that minimizes the arrival time $t_\mathrm{drop}$ of the new request under the constraint that the delay on each currently scheduled requests is smaller than some fraction $\delta$ of the remaining time until the assigned (initially promised) stop time for both pickup and dropoff events. This means customers with a long trip may be delayed proportionally more while customers that were already delayed or that are close to their destination will be delayed less. Secondary objectives for this dispatching algorithm are the minimization of the drive time $t_d = t_\mathrm{drop} - t_\mathrm{pick}$ and the use of the bus with the currently \emph{smallest} occupancy, in order to minimize the impact of the additional delays. We use $\delta = 0.1$ for all simulations on the empirical street networks.

Additional details and results for a third dispatcher algorithm are shown in the Supplementary Information.

\subsubsection*{Networks}
To study the ride-sharing dynamics in qualitatively different network topologies, we use the following model networks with link lengths $l(i,j) = 1$ unless otherwise noted:
\begin{itemize}
    \item[(a)] a minimal network consisting of only two nodes ($N=2$).
    \item[(b)] a star with $N=4$ nodes as a small example of a tree.
    \item[(c)] a Cayley tree (finite Bethe lattice) with $N=94$ nodes.
    \item[(d)] a complete graph with $N=5$ nodes.
    \item[(e)] a ring network (cycle graph) with $N=25$ and $N=100$ nodes.
    \item[(f)] a square lattice with periodic boundaries (torus) with $N=100$ and $N=2500$ nodes.
    \item[(g)] a random geometric network generated from the Delaunay triangulation of $N = 100$ points distributed uniformly at random in the unit square with periodic boundary conditions. The length $l(i,j)$ of the links is given by the Euclidean distance between the connected points.
    \item[(h)] several street networks of various sizes extracted from Open Street Map with the Python package OSMnx \cite{boeing2017osmnx}. For large regions, we discarded the smallest streets according to their tag in the Open Street Map data (see Supplementary Table S1). The length $l(i,j)$ of a link is given by the length of the street it represents. We specifically chose different settings: islands since their street network is self-contained, cities as the typical application area of ride-sharing and rural areas as networks with a qualitatively different structure of the street network. Additional details can be found in the Supplementary Information.
\end{itemize}

\subsubsection*{Request patterns}
All simulation results presented in the main manuscript assume uniformly random requests with $P(i,j) = P_\mathrm{origin}(i) P_\mathrm{dest}(j) = 1/N^2$. Changes in the demand distribution create an effective network topology, affecting the average trip length $\evdel{l}$. It may also cause a change in the distribution of buses on the network, changing the scaling factor $B_{1/2}$, but not affecting the universality. For example, results for additional request distributions on the periodic square lattice are qualitatively identical and are illustrated in the Supplementary Information.

\normalsize

\bibliographystyle{unsrt}
\bibliography{ecobuslit}

\section*{Acknowledgements}
We thank Debsankha Manik, Nils Beyer, Stephan Herminghaus, Jani-Pekka Jokinen, Verena Krall, and Jan Nagler for helpful comments and fruitful discussions. This research was supported by the  German Research Foundation (Deutsche Forschungsgemeinschaft, DFG) through the Cluster of Excellence Center for Advancing Electronics Dresden (cfaed), the European Fund for Regional Development (ERDF/EFRE) through the state of Lower Saxony, and the Max Planck Society.

\section*{Author contributions}
NM, MS and MT conceived and designed research, MS, supported by NM, did the simulations, NM and MS analyzed the simulation results, performed the calculations, and created the figures, advised by MT. NM, MS and MT wrote the manuscript

\section*{Competing Interests}
The authors declare no competing interests.
\newpage
\null
\newpage

\appendix

\renewcommand{\thefigure}{S\arabic{figure}}
\renewcommand{\thetable}{S\arabic{table}}
\renewcommand{\theequation}{S\arabic{equation}}

\setcounter{figure}{0}
\setcounter{table}{0}
\setcounter{equation}{0}

\onecolumngrid

\section*{Supplemental Material}

\subsection*{Ride-sharing model}
In the main manuscript we illustrated our results for ride-sharing dynamics with uniformly random requests [request distribution (i)] and a dispatcher that minimizes the arrival time of requests without allowing any delay for previous requests [dispatcher (A)], both explained below. For simulations on the empirical street networks [Fig.~3(b) in the main manuscript] we applied a dispatcher that allows some delay for already accepted requests [dispatcher (C)] in order to allow small detours in the weighted irregular graphs, in which the shortest path between two nodes is often unique. The results remain qualitatively the same across the different request distributions and dispatcher algorithms (see below).

\subsubsection*{Event based simulation}
We simulate the ride-sharing dynamics with an event-based approach. Events are divided into bus events, where a bus collects or delivers a customer, and request events, where a new request is made. 
Each bus $b\in \{1,...,B\}$ is described by its current location and its route $R_{b}(t) = \left(i_{1}^{(b)},...,i_{n}^{(b)}\right)$ at time $t$ as an ordered list of nodes $i_{k}^{(b)}$ that the bus is scheduled to stop at to pick up or deliver a customer.\\

\textit{Request event: }For a request event at time $t_\mathrm{eq}(k)$ we randomly choose an origin $i(k)$ and a destination $j(k)$ according to a given request distribution $P(i,j)$ (see below). We then find all possible offers of every bus, based on the origin and destination of the request and the current positions and scheduled routes of the buses. We assign the request to the bus $b(k)$ with the ``best'' offer according to the dispatcher algorithm minimizing, for example, the arrival time $t_\mathrm{drop}$ (see below). This bus inserts the two additional stops to pick up and deliver the customer into its route according to the dispatcher decision. Finally, a new request is generated at time $t_\mathrm{req}(k+1) = t_\mathrm{req}(k) + \Delta t$ where $\Delta t$ is distributed exponentially with rate $1/\left<\Delta t\right>$, meaning the requests follow a Poisson process in time.\\

\textit{Bus event: }For a bus event we update the position of the bus. If the bus reaches a scheduled stop, we simply adjust the occupancy of the bus when collecting or delivering a customer (we assume passengers enter and exit buses instantaneously). When a customer is delivered we record the statistics of the trip, such as request time $t_\mathrm{req}$, pickup time $t_\mathrm{pick}$ and drop off time $t_\mathrm{drop}$. Finally, we determine the next event for this bus based on its scheduled route and the bus drives to its next destination. Otherwise, if there are no more scheduled stops, the bus remains at its current location and waits idly until a new request is assigned to it.\\

We start all simulations in the empty state without any requests, all buses are initially idle and uniformly randomly distributed over all nodes in the network. We let the system equilibrate for some time, typically $100$ requests per bus, before measuring for at least $1000$ requests per bus to determine the results reported in the figures.

\subsubsection*{Dispatching algorithms}
In order to confirm the universality of our results we consider different dispatching algorithms that optimize different aspects of the trips. The first two algorithms do not allow for a delay of requests already included in the route. These dispatchers optimize their respective goal functions under the constraint that the times of all other planned stops remain unchanged. As a third example we consider an algorithm that allows for a small delay of already accepted customers proportional to the remaining travel time.

\begin{itemize}
\item[(A)] The most obvious dispatching algorithm optimizes the arrival time $t_\mathrm{drop}$ of the customers. If multiple transporters offer the the same arrival time, this dispatching algorithm chooses the bus with the smallest driving time $t_d = t_\mathrm{drop} - t_\mathrm{pick}$ and then (if still multiple transporters are possible) the bus with the currently largest number of passengers (to potentially let another bus become idle and more effectively serve other requests).
\item[(B)]  The second dispatching algorithm prioritizes minimizing the drive time $t_d = t_\mathrm{drop} - t_\mathrm{pick}$ to minimize the time spent in transit for the customer. Secondary and tertiary criteria are minimizing the arrival time $t_\mathrm{drop}$ and using the bus with the highest occupancy. In effect, this algorithm becomes identical to (A) in the limit of many buses $B \rightarrow \infty$. 
\item[(C)] The last algorithm minimizes the arrival time $t_\mathrm{drop}$ of the new request under the constraint that the delay on each current request is smaller than some fraction $\delta$ compared to the remaining time until the assigned (initially promised) stop time for both pickup and dropoff events. This means customers with a long trip may be delayed proportionally more while customers that were already delayed or that are close to their destination will be delayed less. Secondary objectives for this dispatching algorithm are the minimization of the driving time $t_d = t_\mathrm{drop} - t_\mathrm{pick}$ and the use of the bus with the currently \emph{smallest} occupancy, in order to minimize the impact of the additional delays.

Depending on the value of $\delta$ this has different effects: If $\delta$ is small (but sufficiently large to allow short detours), \emph{only} short detours are allowed and the buses may become more efficient by picking up customers close to their route, especially when fewer buses operate in the system. In the limit $\delta \rightarrow 0$ this algorithm is equivalent to dispatcher (A). If $\delta$ is large, all trips become delayed, as new customers are always inserted before already accepted requests when possible, even to enable inefficient pickups. Customers spend more time waiting and in transit, making the system less efficient overall. We use $\delta = 0.1$ for all simulations on the empirical street networks.
\end{itemize}

\subsubsection*{Networks}
To study the ride-sharing dynamics in qualitatively different network topologies, we use the following model networks:
\begin{itemize}
    \item[(a)] a minimal network consisting of only two nodes ($N=2$) with a distance $l(1,2) = l(2,1) = 1$.
    \item[(b)] a star with $N=4$ nodes as a small example of a tree, where all links have distance $l(i,j) = 1$.
    \item[(c)] a Cayley tree (finite Bethe lattice) with $N=94$ nodes, where all links have distance $l(i,j) = 1$.
    \item[(d)] a complete graph with $N=5$ nodes where all links have distance $l(i,j) = 1$.
    \item[(e)] a ring network (cycle graph) with $N=25$ and $N=100$ nodes, where all links have distance $l(i,j) = 1$.
    \item[(f)] a square lattice with periodic boundaries (torus) with $N=100$ and $N=2500$ nodes, where all links have distance $l(i,j) = 1$.
    \item[(g)] a random geometric network generated from the Delaunay triangulation of $N = 100$ points distributed uniformly at random in the unit square with periodic boundary conditions. The length $l(i,j)$ of the links is given by the Euclidean distance between the connected points.
    \item[(h)] several street networks of various sizes extracted from Open Street Map with the Python package OSMnx \cite{boeing2017osmnx}. For large regions, we discarded the smallest streets according to their tag in the Open Street Map data (see Tab.~\ref{tab.city}). The length $l(i,j)$ of a link is given by the length of the street it represents. We used the largest strongly connected component of these networks without further processing, for example we did not explicitly remove intersections at highways from the possible origin and destination locations.
    We specifically chose different settings: islands since their street network is self-contained, cities as the typical application area of ride-sharing and rural areas as networks with a qualitatively different structure of the street network. In all cases, we assume a fixed velocity $v = 1$ (defining the time scale in the simulation) and a uniform request distribution from all nodes (intersections).
\end{itemize}

\begin{table}[!h]
    \begin{tabular}{| l | l | l | l |}
    \hline
    \textbf{Region} & \textbf{Type} & \textbf{Size} $N$ & \textbf{Streets}\\ \hline
        \hline
    Berlin  & City & 1793 & residential - primary \\ \hline 
    Bornholm & Island &  3969 & all \\ \hline 
    Göttingen & City & 1837 & residential - primary \\ \hline 
    Isle of Man &  Island & 4615 & all \\ \hline 
    Jersey & Island & 1966 & all \\ \hline 
    Korfu & Island & 3662 & all \\ \hline 
    Mallorca & Island & 3599 & tertiary - primary \\ \hline 
    Manhattan & City & 981 & tertiary - primary \\ \hline 
    Modena & City & 5097 & all \\ \hline 
    Ostfriesland & Rural Area & 916 & tertiary - primary \\ \hline 
    Kołobrzeg County & Rural Area & 2127 & all \\ \hline 
    \end{tabular}
    \caption{Empirical street networks. The last column describes the streets used in the network based on their open street map classification ranging from residential (smallest) up to primary (largest) streets or up to highways where applicable.}
    \label{tab.city}
\end{table}

\newpage
\subsubsection*{Request patterns}
All simulation results shown in the main manuscript assume uniformly random requests. Additionally, we also considered different distributions of the request origin and destination in the periodic square lattice (torus) network. In all cases, the requests are drawn from a distribution $P(i,j)$ such that the origin $i$ and the destination $j$ are chosen independently, $P(i,j) = P_\mathrm{origin}(i) P_\mathrm{dest}(j)$.
\begin{itemize}
\item[(i)] In the simplest case, we assume symmetric and uniformly distributed origins and destinations $P_\mathrm{origin}(i) = P_\mathrm{dest}(i) = 1/N$.

\item[(ii)] We also consider symmetric, but random origin and destination probabilities. We draw random variables $X_i \sim \mathrm{Uniform}[0,1]$ and then normalize them to define the origin and destination probabilities $P_\mathrm{origin}(i) = P_\mathrm{dest}(i) = \frac{X_i}{\sum_{i=1}^N X_i}$. 

\item[(iii)] As before, we consider random origin and destination probabilities, but now draw them independently for origin and destination. We draw random variables $X_i \sim \mathrm{Uniform}[0,1]$ and $Y_i \sim \mathrm{Uniform}[0,1]$ then normalize them to define the origin and destination probabilities $P_\mathrm{origin}(i) = \frac{X_i}{\sum_{i=1}^N X_i}$ and $P_\mathrm{dest}(i) = \frac{Y_i}{\sum_{i=1}^N Y_i}$.


\item[(iv)] To more closely represent the spatial distribution of requests in a city we use a unimodal distribution of origin nodes on the torus. We choose a center $i^*=0$. Each node is then assigned a weight $X_i = \mathrm{exp}\left( -\frac{2 l_{i^*i}^2}{ \mathrm{max}_{j}(l_{i^*j})^2 } \right)$, which corresponds to a two-dimensional Gaussian distribution with mean $0$ and standard deviation $\mathrm{max}_{j}(l_{i^*j}) / 2$. To obtain the probabilities we finally normalize these weights as above, $P_\mathrm{origin}(i) = P_\mathrm{dest}(i) = \frac{X_i}{\sum_{i=1}^N X_i}$.

\item[(v)] We also consider an asymmetric version of (iv). We choose two centers $i^*=0$ (for the origin) and $j^*$ (for the destination), where $j^*$ is chosen such that the distance to $i^*$ is maximal. Each node is then assigned weights $X_i = \mathrm{exp}\left( -\frac{2 l_{i^*i}^2}{ \mathrm{max}_{j}(l_{i^*j})^2 } \right)$and $Y_i = \mathrm{exp}\left( -\frac{2 l_{j^*i}^2}{ \mathrm{max}_{k}(l_{j^*k})^2 } \right)$. We then normalize to obtain $P_\mathrm{origin}(i) = \frac{X_i}{\sum_{i=1}^N X_i}$ and $P_\mathrm{dest}(i) = \frac{Y_i}{\sum_{i=1}^N Y_i}$. 

\end{itemize}

\clearpage

\subsection*{Efficiency and susceptibility at high load}
In the main manuscript, we define the efficiency of a ride-sharing system via the scaling of the number of scheduled customers as the normalized load $x$ increases. The efficiency $E = \lim_{x\rightarrow\infty} \left(\frac{\evdel{C}}{x}\right)^{-1}$ measures how well all requests are served, the susceptibility $\chi = \left(\frac{\mathrm{d}\,\evdel{C}}{\mathrm{d}\,x}\right)^{-1}$ measures how well additional requests are served.
Using the relation of $\evdel{C}$ to the waiting and drive time derived in the main manuscript $\evdel{C} = \frac{v\,x}{2\,\evdel{l}}\left( \evdel{t_w} + \evdel{t_d} \right)$ [Eq.~(3)], we write the efficiency as
\begin{equation}
    E = \lim_{x\rightarrow\infty} \left(\frac{\evdel{C}}{x}\right)^{-1} = \left(\frac{v}{\evdel{l}}\left( \evdel{t_w} + \evdel{t_d} \right)\right)^{-1}
\end{equation}
and the susceptibility as
\begin{eqnarray}
    \chi &=& \left(\frac{\mathrm{d}\,\evdel{C}}{\mathrm{d}\,x}\right)^{-1} \\
      &=& \left(\frac{v}{\evdel{l}}\left( \evdel{t_w} + \evdel{t_d} \right) + \frac{v\,x}{\evdel{l}} \frac{\mathrm{d}\,\left(\evdel{t_w} + \evdel{t_d}\right)}{\mathrm{d}\,x} \right)^{-1}
\end{eqnarray}
In the limit of high load, when the request rate is sufficiently large, additional requests become identical to requests already in the system. These requests then do not affect the average waiting and drive time and these times become constant. Thus, in this limit, $x \rightarrow \infty$, the second term vanishes and both definitions become identical
\begin{equation}
    \lim_{x \rightarrow \infty} \chi = E = \left(\frac{v}{\evdel{l}}\left( \evdel{t_w} + \evdel{t_d} \right)\right)^{-1} \,.
\end{equation}

\newpage

\subsection*{Topological factors of empirical street networks}

Table \ref{tab.city_fit} shows the scaling factors $E_\mathrm{max}$ and $B_{1/2}$ for the empirical street networks in Fig.~3(b) in the main manuscript. The factors were determined by fitting the universal curve $E = E_\mathrm{max}\frac{B}{B + B_{1/2}}$ to the simulation results. The maximum possible efficiency $E_\mathrm{max}$ is approximately similar for all networks, confirming our expectation that it should predominantly depend on the dispatcher algorithm, not on the network topology. The half-efficiency fleet size varies strongly across topologies. We note in particular that the low values for Mallorca, Manhattan and Ostfriesland are likely due to discarding the smallest streets and thus having proportionally fewer dead ends in the network.

\begin{table}[h]
    \begin{tabular}{| l | l | l | l |}
    \hline
    \textbf{Region} & $B_{1/2}$ & $E_\mathrm{max}$ \\ \hline
        \hline
    Berlin & 325 $\pm$ 40 & 0.79 $\pm$ 0.03 \\ \hline 
    Bornholm & 201 $\pm$ 20  &  0.80 $\pm$ 0.03  \\ \hline 
    Göttingen & 391 $\pm$ 10  &  0.71 $\pm$ 0.02 \\ \hline 
    Isle of Man & 440 $\pm$ 90  &  0.79 $\pm$ 0.08 \\ \hline 
    Jersey & 240 $\pm$ 10 & 0.76 $\pm$ 0.02\\ \hline 
    Korfu &  222 $\pm$ 30 &   0.75 $\pm$ 0.04\\  \hline 
    Mallorca &  114 $\pm$ 3 &   0.78 $\pm$ 0.01 \\  \hline 
    Manhattan &  80 $\pm$ 6 &  0.82 $\pm$ 0.02 \\  \hline 
    Modena & 1200 $\pm$ 190 & 0.67 $\pm$ 0.05 \\  \hline 
    Ostfriesland & 110 $\pm$ 4 &  0.79 $\pm$ 0.01 \\  \hline 
    Kołobrzeg County & 300 $\pm$ 20 & 0.73 $\pm$ 0.02 \\ \hline 
    \end{tabular}
    \caption{\textbf{Scaling parameters for the empirical street networks.} $B_{1/2}$ encodes the effect of the network size and topology. $E_\mathrm{max}$ is approximately constant across all networks and describes maximum possible efficiency of the dispatcher algorithm. See also Tab.~\ref{tab.city}.} 
    \label{tab.city_fit}
\end{table}

\clearpage

\subsection*{Mean-field approximation of $B_{1/2}$}

For a range of model networks, we can exploit symmetries to calculate the scaling factor $B_{1/2}$. We use the asymptotic efficiency curve in the perfect service limit [compare Eq.~(4) in the main manuscript] to compute the scaling factor using the scaling of the average waiting time under mean-field conditions. We first consider the number of buses on a link and calculate the expected waiting times for the different trips. Averaging over all trips then gives the average waiting time $\evdel{t_w}$. In the following we give four examples for these calculations. All values of $B_{1/2}$ are reported in Tab.~\ref{tab.gamma} below.

\subsubsection*{Minimal two node graph}
In the minimal graph, there are only two directed links, $M = 2$. The buses $B$ are distributed equally on both links with $B/2$ buses per link. Since both nodes are symmetric, we consider only trips originating from node $1$ (waiting times for trips from node $2$ are identical by symmetry). These requests either go to node $1$ with waiting time $\evdel{t_w^{11}}=\evdel{t_w^{22}}$ or go to node $2$ with waiting time $\evdel{t_w^{12}} = \evdel{t_w^{21}}$.

All $B/2$ buses that are driving on the link to node $1$ are eligible to accept a request to node $1$. These buses are distributed on the link of length $l_{21} = 1$, that means with expected distance $1/(B/2) = 2/B$. The expected waiting time for a customer then corresponds to half of this distance (in the best case a bus is there immediately, in the worst case a bus is $2/B$ away), i.e. 
\begin{equation}
    \evdel{t_w^{11}} = \frac{1}{2} \frac{2}{vB} = \frac{1}{vB}. 
\end{equation}
Since all incoming buses must continue on to node $2$, the same calculation gives the waiting time \begin{equation}
    \evdel{t_w^{12}}= \frac{1}{vB}.
\end{equation}

Averaging over all trips gives 
\begin{equation}
    \evdel{t_w} = \frac{1}{2} \left(\evdel{t_w^{11}}+\evdel{t_w^{12}}\right) = \frac{1}{vB} = B^\mathrm{mini}_{1/2} \frac{\evdel{l}}{vB}.
\end{equation}
Together with $\evdel{l} = 1/2$ and Eq.~(9) in the main manuscript, we find
\begin{equation}
    B^\mathrm{mini}_{1/2} = 2 \,.
\end{equation}

\subsubsection*{Complete graph}
In the complete graph all buses are distributed on the $M = N(N-1)$ directed links. Again, all nodes are identical under symmetry and we consider only trips starting from node $i$. For trips from a node to itself, all $(N-1) B/M$ incoming buses are eligible, resulting in 
\begin{equation}
    \evdel{t_w^{ii}} = \frac{1}{2} \frac{M}{v (N-1) B} = \frac{N}{2vB}. 
\end{equation}
For trips to a specific other node, only the $B/M$ buses going to this node can serve the request, giving 
\begin{equation}
    \evdel{t_w^{io}} = \frac{1}{2} \frac{M}{v B} = \frac{N(N-1)}{2vB}.
\end{equation} 
Averaging these waiting times over all trips gives \begin{equation}
    \evdel{t_w} = \frac{N^2/2 - N + 1}{v B} = B^\mathrm{comp}_{1/2} \frac{\evdel{l}}{vB}
\end{equation}
and 
\begin{equation}
    B^\mathrm{comp}_{1/2} = 
    \frac{N^2/2 - N + 1}{ N(N-1) / N^2} = \frac{N^3 - 2N^2 + 2N}{2N - 2}.\nonumber
\end{equation} 
In the limit of large networks $N \rightarrow \infty$ we have 
$\evdel{l} \rightarrow 1$ such that $B^\mathrm{comp}_{1/2} = \frac{N^2}{2}$. For the $N=5$ complete graph we find $B^\mathrm{comp}_{1/2} = 85/8 = 10.625$.


\subsubsection*{Ring}
In a ring, $B$ buses are distributed over $M = 2N$ directed links with $B/(2N)$ buses per link. Again, all nodes are identical under symmetry and we consider only trips starting from node $i$. We assume that buses go around the ring in a fixed direction, half going clockwise, half counter-clockwise, as they always have new customers when the normalized request rate $x$ is large. Under this constraint, all buses incoming from $i-1$ to $i$ continue to $i+1$ and vice versa. For trips from $i$ to $i$, buses on both incoming links are eligible and we get the waiting time 
\begin{equation}
\evdel{t_w^{ii}} = \frac{1}{2} \frac{2N}{2vB} = \frac{N}{2vB}\,.
\end{equation}
For all other trips, only buses from one direction are eligible with \begin{equation}
\evdel{t_w^{ii'}} = \frac{N}{vB} \,,
\end{equation}
except for the node directly opposite to $i$ when $N$ is even. Averaging over all trips gives
\begin{equation}
    \evdel{t_w} = \frac{2N^2 - N}{4N^2} \frac{M}{vB} = B^\mathrm{ring}_{1/2} \frac{\evdel{l}}{vB}.
\end{equation}
For large $N$ we find to leading order 
\begin{equation}
B^\mathrm{ring}_{1/2} = 4  \,.  
\end{equation}

\subsubsection*{Star}
A star graph is a tree graph with a single node in the center directly connected to all other nodes via $M = 2 (N - 1)$ directed links. In this case, not all links are identical. There are two types of links, those going outwards from the center and those going inwards to the center. However, since each bus that goes out from the center must come back, there are again the same number $B/M$ buses on each directed link. We now have to distinguish all possible different cases for a request: from inner node to inner node $(i,i)$, from inner node to outside node $(i,o)$, from an outside node to the inner node $(o,i)$, from an outside node to another outside node $(o,o')$ and from an outside node to itself $(o,o)$, where all outside nodes are identical under symmetry.

Let us first consider requests originating at the inner node: $(i,i)$-requests have to wait for any of the $(N-1) B / M$ buses inbound to the inner node, they thus have an average waiting time of
\begin{equation}
    \evdel{t_w^{(i,i)}} = \frac{M}{2 B (N-1) v} = \frac{1}{vB}\,.    
\end{equation}

$(i,o)$-requests have to wait for the buses that go to their outside node, that means a fraction $1/(N-1)$ out of the total $(N-1) B / M$ buses. Consequently, they have an average waiting time of
\begin{equation}
    \evdel{t_w^{(i,o)}} = \frac{M}{2 vB} = \frac{N-1}{vB}\,.    
\end{equation}

$(o,i)$-requests have to wait for any of the buses that are inbound to their node since all of them have to travel back through the inner node. The same holds for $(o,o)$-requests. This means they are waiting for one out of $B / M$ buses and have an average waiting time of
\begin{equation}
    \evdel{t_w^{(o,i)}} = \evdel{t_w^{(o,o)}} = \frac{M}{2 vB} = \frac{N-1}{vB}\,.    
\end{equation}

$(o,o')$-requests have to wait for the buses that go to their specific destination node $o'$, that means a fraction $1/(N-2)$ out of the total $B / M$ buses inbound at $o$. They have an average waiting time of
\begin{equation}
    \evdel{t_w^{(o,o')}} = \frac{M (N-2)}{2 vB}  = \frac{(N-1)(N-2)}{vB}\,.    
\end{equation}


Averaging these waiting times, weighted with the occurrence probabilities of the corresponding request type, we obtain
\begin{eqnarray}
    \evdel{t_w} &=& \frac{1}{N^2} \evdel{t_w^{(i,i)}} + \frac{N-1}{N^2} \evdel{t_w^{(i,o)}} + \frac{N-1}{N^2} \evdel{t_w^{(o,i)}} \nonumber\\
    && + \frac{N-1}{N^2} \evdel{t_w^{(o,o)}} + \frac{(N-1)(N-2)}{N^2} \evdel{t_w^{(o,o')}} \nonumber\\
     &=& \frac{(N-1) \left(N^4-6N^3+16N^2-18N+8\right)}{v B N^2 \left(N - 1\right)} \nonumber\\ 
     &=& \frac{N^4-6N^3+16N^2-18N+8}{N^2 \, \evdel{l}} \frac{\evdel{l}}{vB} \approx \frac{N^2}{\evdel{l}} \frac{\evdel{l}}{vB}\,, \nonumber
\end{eqnarray}
where the last expression holds in the limit of large networks, $N \rightarrow \infty$. In this limit we find $B^\mathrm{star}_{1/2} = \frac{N^2}{2}$. For the $N=4$-node star we have $\evdel{l} = 18/16$, resulting in $B^\mathrm{star}_{1/2} = 32/9 \approx 3.56$.


\begin{table}[!h]
    \begin{tabular}{| l | l | l | l |}
    \hline
    \textbf{Network} & \textbf{Size} $N$ & $B_{1/2}$ (theory) & $B_{1/2}$ (fit) \\  \hline
        \hline
    Minimal & 2 & 2 & 2.03 $\pm$ 0.01 \\ \hline             
    Ring & 25 & 4 & 4.97 $\pm$ 0.1 \\ \hline       
    Ring & 100 & 4 & 5.12 $\pm$ 0.1 \\ \hline      
    Star & 4 & 3.56 & 4.4 $\pm$ 0.4 \\ \hline 
    Cayley tree & 94 & - & 540 $\pm$ 20 \\ \hline        
    Complete Graph & 5 & 10.625 & 12.8 $\pm$ 0.3  \\ \hline  
    Torus & 100  & - & 176 $\pm$ 5 \\ \hline    
    Torus & 2500  & - & 2760 $\pm$ 30 \\ \hline    
    Random geometric & 100 & - & 360 $\pm$ 20 \\ \hline 
    \end{tabular}
    \caption{\textbf{Theoretical and numerical scaling factor for artificial networks.} Numerical scaling factors are determined by fitting the simulations results for $B \ge 600$ ($B \ge 7500$ for the torus with $N=2500$ nodes) to the asymptotic efficiency $E = \left(1 + B_{1/2} / B\right)^{-1}$ for large $B$. 
    }
    \label{tab.gamma}
    \vspace{-0.5cm}
\end{table}

\clearpage

\subsection*{Scaling of the efficiency close to the perfect service limit}
\begin{figure*}[h]
        \centering
	\includegraphics[width=0.95\textwidth]{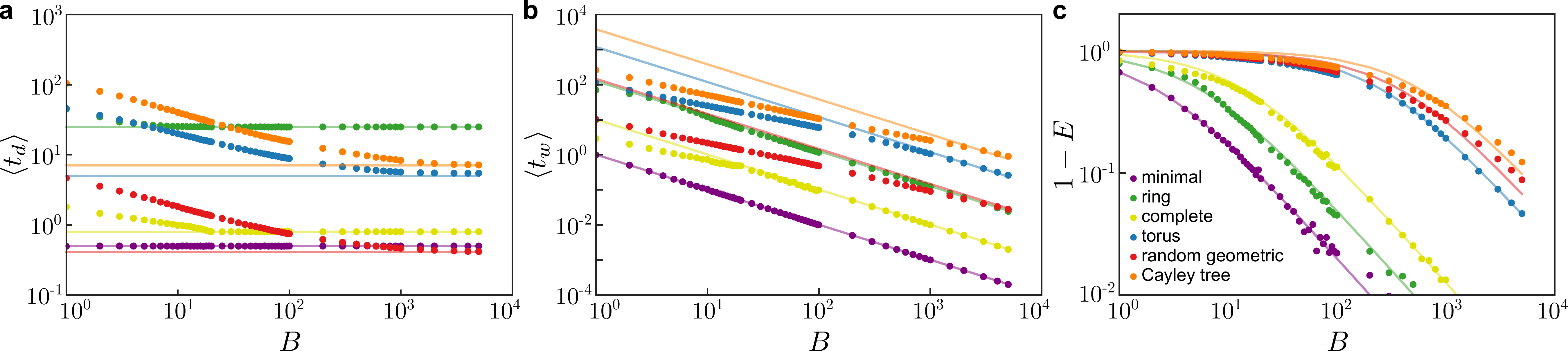}
        \caption{\textbf{Asymptotic scaling of ride-sharing efficiency.} Scaling of the drive time $\evdel{t_d}$ and the wait time $\evdel{t_w}$ time with the number of buses and the resulting asymptotic efficiency curves. (a) The drive time $\evdel{t_d}$ quickly approaches its minimum possible value, $\evdel{t_d} \rightarrow \evdel{l} / v \sim B^0$ for large $B$.
        (b) The waiting time $\evdel{t_w}$ goes to zero as $\evdel{t_w} \approx \gamma \, \left(\evdel{l}/v\right) B^{-1}$ in the limit of many buses.
        (c) The efficiency $E$ [Eq.~(1) in the main manuscript, evaluated at $x=7.5$] for the different topologies is well approximated by the prediction [Eq.~(\ref{eq:derived_universal_SI})] using only the leading order scaling of the waiting time and drive time [compare Fig.~2(d) in the main manuscript].
        }
        \label{fig:time_scaling}
\end{figure*}

Figure~\ref{fig:time_scaling} illustrates the leading order scaling of the driving time $\evdel{t_d} \propto B^0$ and waiting time $\evdel{t_w} \propto B^{-1}$ close to the perfect service limit with many buses, $B \rightarrow \infty$. Using only this leading order scaling, we express the efficiency $E$ as 
\begin{equation}
    E \sim \frac{1}{1 + \left(\frac{B}{B_{1/2}}\right)^{-1}} \,. \label{eq:derived_universal_SI}
\end{equation}
This scaling is well confirmed by the simulations [Fig.~\ref{fig:time_scaling}(c)].

\clearpage

\subsection*{Robustness of results}

\begin{figure*}[!h]
        \centering
	\includegraphics[width=0.95\textwidth]{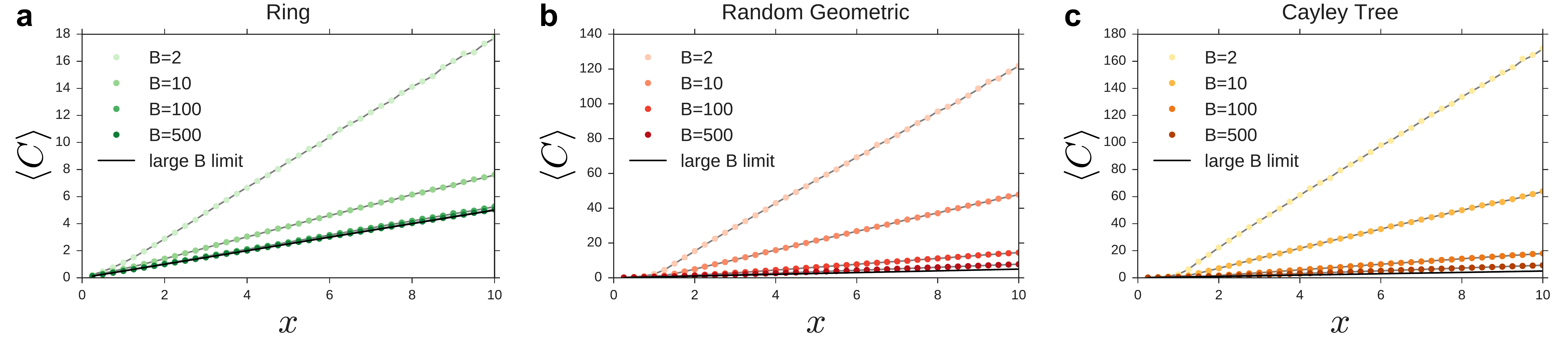}
        \caption{\textbf{Convergence to optimal ride-sharing in different network topologies.} The scaling of the number of scheduled customers $\evdel{C}$ converges to the optimal service limit $\evdel{C} = x$ (black line) for different topologies [compare Fig.~2(c) in the main manuscript].}
        \label{fig.supp_b_scaling}
\end{figure*}

The efficiency of ride-sharing changes with the number of buses and the request rate. In the main manuscript we illustrated the convergence of $\evdel{C}$ to optimal service, $\evdel{C} = x$, in the torus network. We observe the same for the other topologies, see Fig.~\ref{fig.supp_b_scaling}. The convergence is faster for some networks, such as the ring, signifying better conditions for ride-sharing already for smaller $B$. The scaling factor $B_{1/2}$, describing the number of buses required to reach half efficiency, is correspondingly smaller for the ring than for the other networks.

The results presented in the main manuscript illustrate the universality of the ride-sharing efficiency for uniformly random requests (i) and arrival time minimizing dispatcher algorithm (A). They remain qualitatively unchanged for other request distributions (i-v) and dispatcher algorithms (A-C).

\begin{figure*}[!h]
        \centering
	\includegraphics[width=0.8\textwidth]{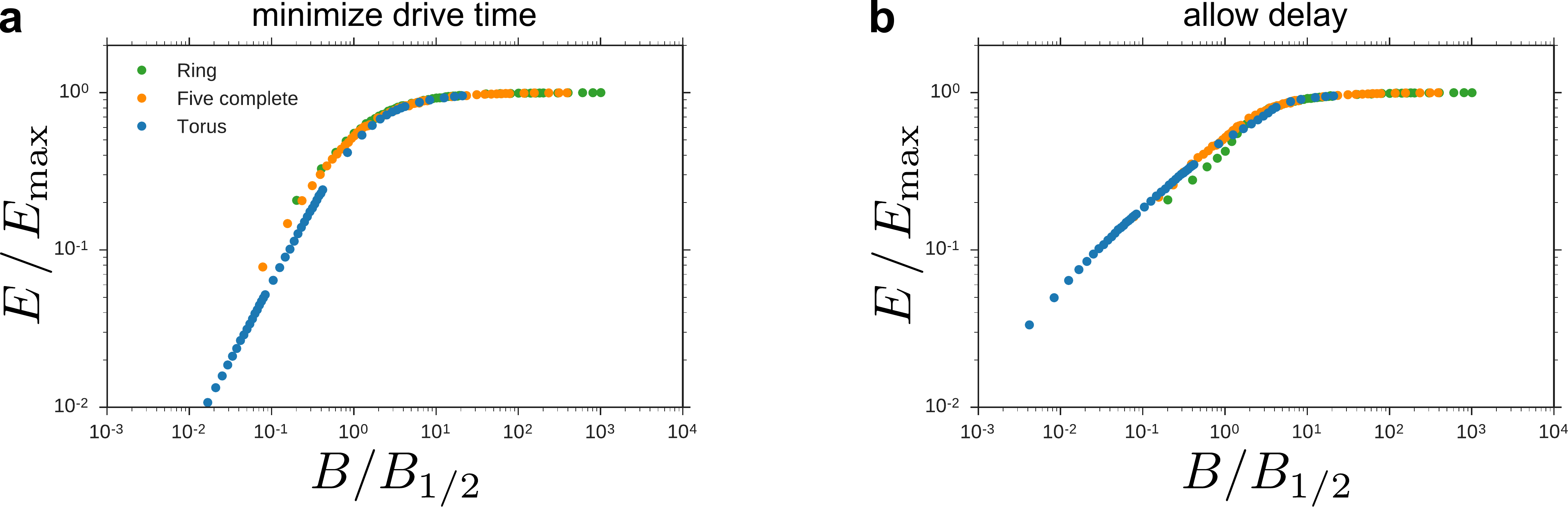}
        \caption{\textbf{Ride-sharing universality with respect to dispatcher algorithms.} Different dispatcher algorithms [dispatcher (B), left] and [dispatcher (C), right] give rise to different universal curves, each with specific $B_{1/2}$ and $E_\mathrm{max}$ depending on topology, request pattern and dispatcher. Here, $E_\mathrm{max} \approx 1$ in all cases.}
        \label{fig.dispatcher_algorithms}
\end{figure*}

\begin{figure*}[!h]
        \centering
	\includegraphics[width=0.4\textwidth]{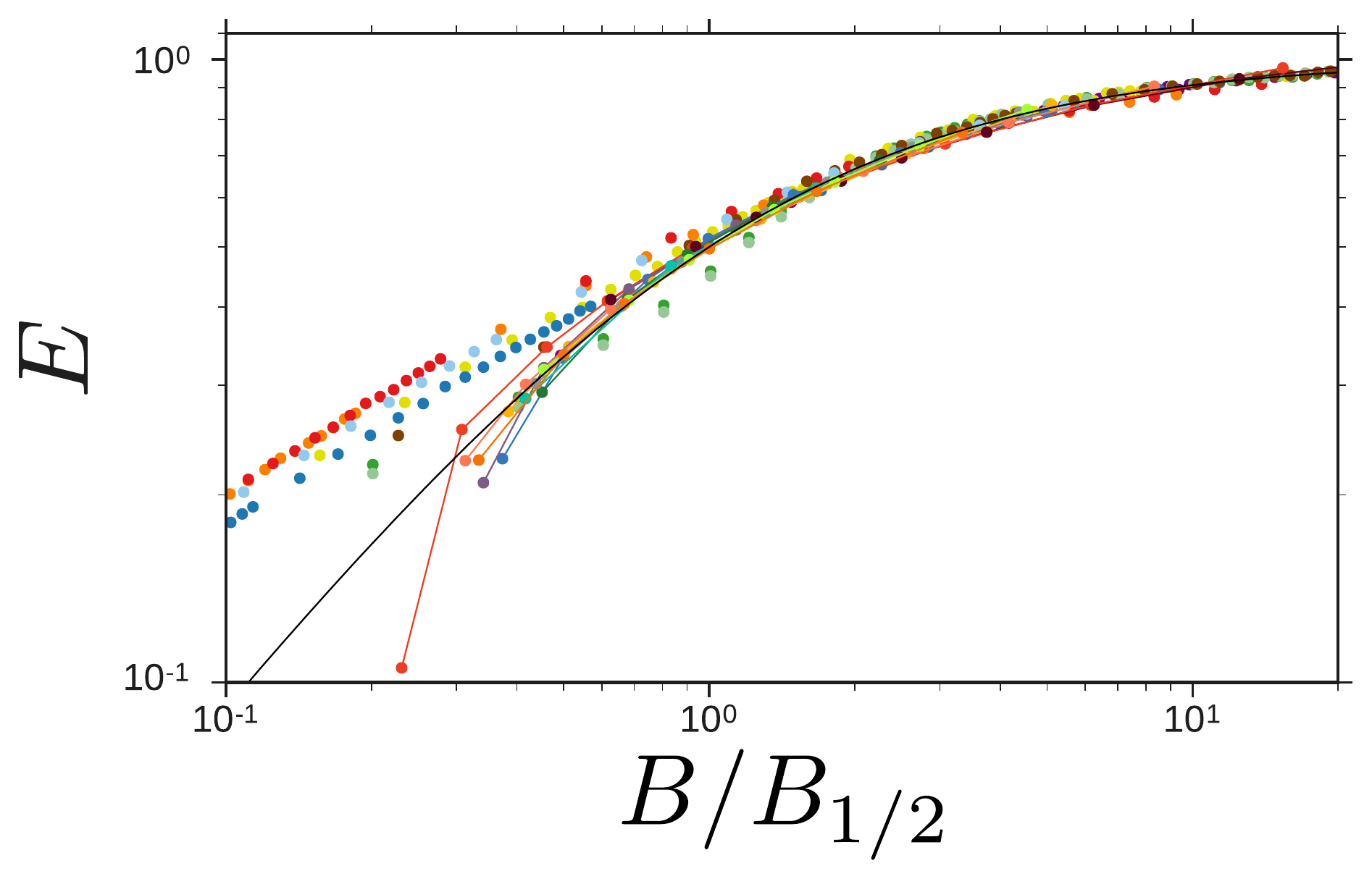}
        \caption{\textbf{Asymptotic universality holds across dispatcher algorithms.} Different dispatcher algorithms [dispatcher (A) for model networks (points), dispatcher (C) for real networks (points and lines)] follow different universal functions $f(\cdot)$. The asymptotic scaling for large $B$ (black line, Eq.~(7) in the main manuscript) holds for both dispatcher since both have an asymptotically constant driving time and a waiting time scaling as $B^{-1}$ (compare Fig.~3 in the main manuscript).}
        \label{fig.dispatcher_algorithms_comparison}
\end{figure*}

Different dispatcher algorithms naturally behave differently than assumed in the derivation in the main manuscript. While most dispatchers will attempt to keep the drive time small to ensure that customers arrive at their destination, in some cases certain assumptions necessary for the arguments in the main text are not valid. In particular, when allowing delays of the trips, $\evdel{t_d}$ may not approach $\evdel{l}/v$ [dispatcher (C)]. Nonetheless, the general scaling form [Eq.~(2) in the main text],
\begin{eqnarray}
    E(B) \sim E_\mathrm{max} \, f\left( \frac{B}{B_{1/2}}\right) \, \nonumber
\end{eqnarray}
remains valid. The efficiency curves are universal for each dispatcher algorithm, illustrated in Fig.~\ref{fig.dispatcher_algorithms}. As long as the driving time becomes constant in the perfect service limit and the waiting time decays as $B^{-1}$, the efficiency curve for large $B$ follows the asymptotic curve derived in the main manuscript (Eq.~7 in the main manuscript, see Fig.~\ref{fig.dispatcher_algorithms_comparison}) with appropriate $B_{1/2}$ and $E_\mathrm{max}$.

\begin{figure}[!h]
        \centering
	\includegraphics[width=0.4\textwidth]{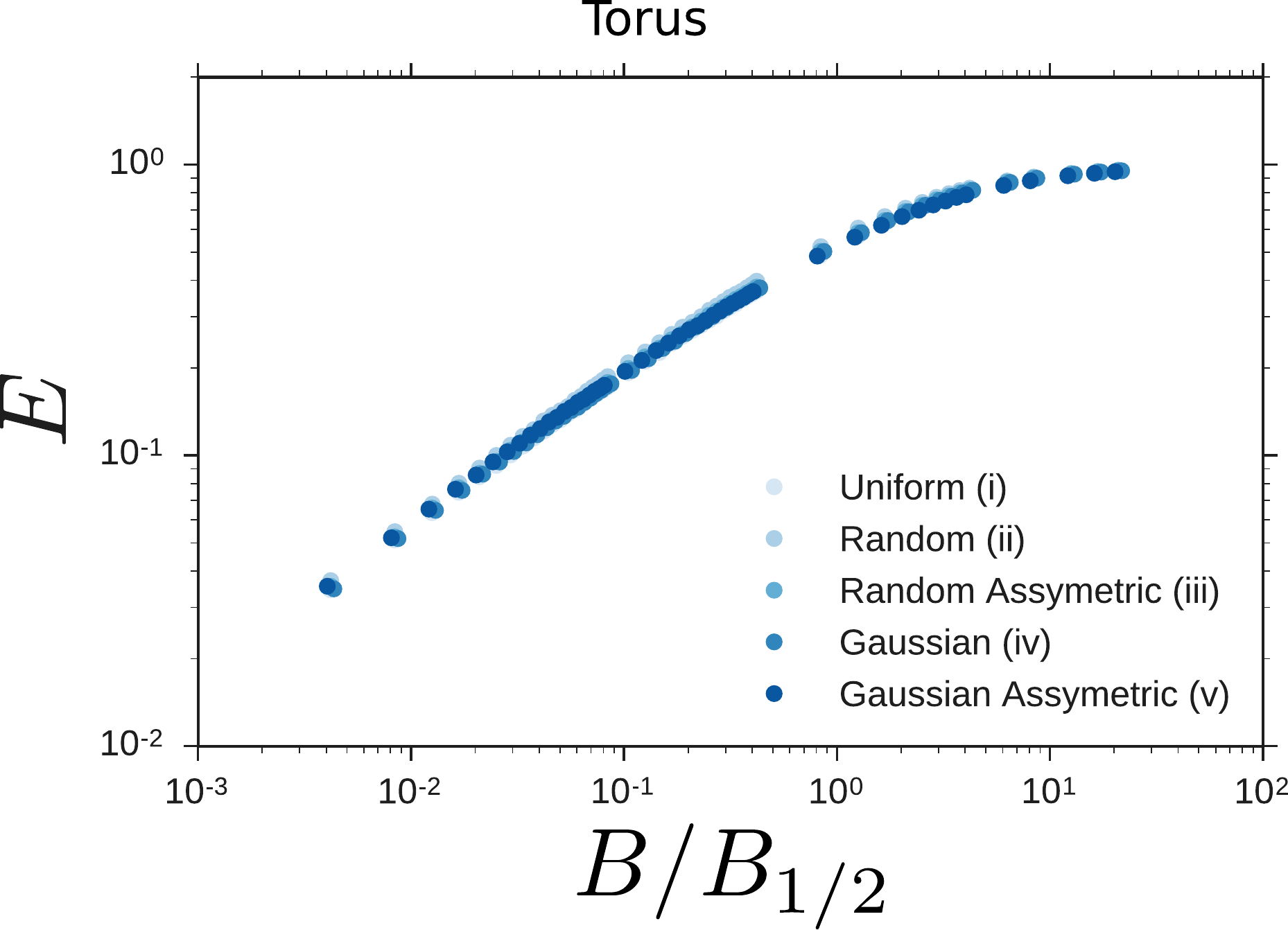}
        \caption{\textbf{Ride-sharing universality with respect to request distributions.} Different request distributions (i - v) do not influence the universality of the ride sharing efficiency on the torus with $N=121$ nodes. The different request distributions change the effective topology of the network, modifying the average trip length $\evdel{l}$ and the distribution of the buses on the network. Using the correctly weighted $\evdel{l}$ and the appropriate rescaling factor $B_{1/2}$, all efficiency curves collapse to the same curve.}
        \label{fig.request_patterns}
\end{figure}

Additionally, we defined different request distributions on a torus with $N = 121$ nodes, modelling uniform, random and partially directed requests. In all cases we find almost identical results (Fig.~\ref{fig.request_patterns}).

\end{document}